\documentclass[a4paper,12pt,aip,jcp,reprint]{revtex4-1}
\usepackage{color}
\usepackage{graphicx}
\usepackage{enumerate}
\usepackage{amsfonts,amssymb}
\usepackage{amsmath}
\usepackage{stmaryrd}
\usepackage{dcolumn}
\usepackage{bm}
\usepackage{bbm}
\usepackage{algpseudocode}
\usepackage{algorithm}
\usepackage[all,cmtip]{xy}
\usepackage{url}
\usepackage{braket}
\usepackage[version=3]{mhchem} 
\usepackage{subcaption}
\begin{document}

\title[]{An efficient stochastic algorithm for
the perturbative density matrix renormalization group in large active spaces}
\author{Sheng Guo}
\author{Zhendong Li}
\author{Garnet Kin-Lic Chan}\email{gkc1000@gmail.com}
\affiliation{Division of Chemistry and Chemical Engineering,
California Institute of Technology, Pasadena, CA 91125, USA}
\begin{abstract}
  We present an efficient stochastic algorithm for the recently introduced perturbative density matrix renormalization group (p-DMRG)  method
  for large active spaces. The stochastic implementation
bypasses the computational bottleneck involved in solving the first order equation in the earlier
deterministic algorithm. 
We demonstrate the efficiency and accuracy of the algorithm on the \ce{C2} and \ce{Cr2} molecular benchmark systems.
 \end{abstract}

\maketitle

Molecular electronic structure with a mix of static and dynamic correlation remains a challenging aspect of quantum chemistry.
It is often treated within a complete active space (CAS) approach
for the static correlation plus a second-order perturbative treatment of the dynamic correlation,
such as with CASPT2\cite{andersson_second-order_1990, roos_multiconfigurational_1996} (complete-active-space second-order perturbation theory) and NEVPT2\cite{angeli_introduction_2001,angeli_n-electron_2001, angeli_n-electron_2002} ($N$-electron valence state perturbation theory).
However, in many cases the second-order treatment of correlations outside of the active space is not quantitatively accurate,
and it becomes necessary to enlarge the active space with  orbitals of intermediate correlation strength.
For instance, in $3d$ transition metal complexes, the virtual $4d$, semi-core ($3s$ and $3p$), or valence ligand orbitals often need
to be included in the active space for good accuracy\cite{ANDERSSON1992507,kurashige_second-order_2011, guo_n-electron_2016}.
The resulting very large number of active orbitals with a mix of different
correlation character renders the density matrix renormalization group (DMRG)\cite{white_density_1992,white_density-matrix_1993, white_ab_1999,mitrushenkov2001quantum,chan_highly_2002,legeza2003controlling,sharma_spin-adapted_2012,roberto,keller_efficient_2015,yanai_density_2015,chan_matrix_2016}, whose strength lies
in treating large active spaces where all the orbitals are strongly interacting, relatively inefficient,
and the subsequent dynamic correlation treatment also becomes very difficult
\cite{kurashige_second-order_2011,sharma_communication:_2014, Naoki_caspt2, guo_n-electron_2016, sharma2017combining, Alex_t_nevpt, Reigher_dmrg_nevptpt, Alex_t_dmrg_pt}.

Recently, we developed a perturbative DMRG (p-DMRG)\cite{p_DMRG} algorithm to efficiently target near-exact energies in
large orbital spaces where there is a mix of strongly and intermediately correlated orbitals.
We showed that the p-DMRG  provides, in practice, benchmark quality energies as accurate as those obtained by
variational DMRG calculations, but with a significantly reduced computational cost.
The essential idea is to treat the strong correlation with a variational DMRG
calculation with small bond dimension $M_0$ (which can be systematically increased) within the full
orbital space, and the residual correlation by second-order perturbation theory (PT2) within the same orbital space.
The total energy obtained by p-DMRG converges very quickly with $M_0$, which can thus be chosen significantly
smaller than the bond dimension $M$ in the fully variational calculation.
The basic idea of p-DMRG is similar in spirit to that of  selected configuration interaction (SCI) plus
perturbation theory\cite{huron1973iterative,buenker1974individualized,harrison1991approximating, schriber2016communication,tubman2016deterministic,liu2016ici,holmes_heat-bath_2016,sharma2017semistochastic,garniron2017hybrid},
and in particular, the recently developed Heat-bath CI+PT (HCI+PT)\cite{holmes_heat-bath_2016,sharma2017semistochastic},
where the static correlation is treated by configuration interaction within a selected set of important determinants
defined by a threshold, which can be tuned to systematically increase the accuracy much like $M_0$ in p-DMRG, with the residual
dynamic correlation in the same orbital space treated by standard PT2.

In this Communication, we describe an efficient stochastic algorithm that further speeds up the p-DMRG algorithm
by avoiding having to solve the first order equation deterministically. Such a stochastic extension to HCI+PT
has already been described~\cite{sharma2017semistochastic}.
However, in the selected CI setting, the stochastic extension is very natural due to a clear
separation of the variational and first order determinant spaces. Since in p-DMRG,
the zeroth order wavefunction is a matrix product state (MPS)\cite{chan_matrix_2016}, the stochastic extension
is less straightforward and this Communication is thus concerned primarily with its correct and efficient formulation.

To begin with, we assume that a zeroth order wavefunction $\ket{\Psi^{(0)}}$,
expressed as an MPS with a small bond dimension $M_0$, has been optimized by the standard variational DMRG algorithm.
Given a partitioning of the Hamiltonian, $\hat{H}=\hat{H}_0+\hat{V}$ with $\hat{H}_0\ket{\Psi^{(0)}}=E_0\ket{\Psi^{(0)}}$, the first order wave function $\ket{\Psi^{(1)}}$ is defined by the first order equation,
\begin{equation}
  Q(\hat{H}_0-E_0)Q\ket{\Psi^{(1)}}=-Q\hat{V}\ket{\Psi^{(0)}},\label{eq:1st}
\end{equation}
where $Q = 1- P$ and $P = \ket{\Psi^{(0)}}\bra{\Psi^{(0)}}$ are projectors.
Once $\ket{\Psi^{(1)}}$ is obtained from Eq. \eqref{eq:1st}, which
in our implementation is achieved by minimizing the Hylleraas functional using the MPSPT algorithm\cite{sharma_communication:_2014},
the second order energy $E_2=\bra{\Psi^{(1)}}Q\hat{V}\ket{\Psi^{(0)}}$ can be computed.
For very large active spaces with 50-100 orbitals, the required bond dimension
$M_1$ for expanding $\ket{\Psi^{(1)}}$, which typically scales like $O(K^2M_0)$,
can be quite large ($\gg10^4$). This creates a significant cost both in computation
and in storage. In the previous deterministic p-DMRG algorithm\cite{p_DMRG}, we used a sum of MPS representation
 $\ket{\Psi^{(1)}}=\sum_i \ket{\Psi^{(1)}_i}$  as well as
extrapolation, which helped alleviate the computational cost.
In contrast, a stochastic algorithm can be expected to essentially eliminate this bottleneck, at the cost
of introducing statistical errors.

To develop a stochastic variant of p-DMRG, we first rewrite $E_2$ as
\begin{eqnarray}
E_2=-\bra{\Psi^{(0)}}\hat{V}Q[Q(\hat{H}_0-E_0)Q]^{-1}Q\hat{V}\ket{\Psi^{(0)}},\label{eq:e2init}
\end{eqnarray}
and then aim to find an explicit expression for $E_2$ as a sum over
terms, ideally of a similar form to $\sum_{i}-\frac{|\langle D_i|\hat{V}|\Psi^{(0)}\rangle|^2}{E_i-E_0}$ as appears in HCI+PT
with $|D_i\rangle$ being a determinant, which can then be sampled stochastically.
In our previous paper\cite{p_DMRG},
we chose $\hat{H}_0$ as
\begin{equation}
  \hat{H}_0 = P E_0 P + Q\hat{H}_d Q,\label{eq:H0}
\end{equation}
where $\hat{H}_d$ contains those operators in $\hat{H}$
which do not change the occupation numbers of spatial orbitals, i.e.,
\begin{equation}
  \hat{H}_d = \sum_{p} h_{pp} \hat{E}_{pp} + \frac{1}{2}\sum_{pq} (pp|qq)\hat{e}_{pqqp} +
  \frac{1}{2}\sum_{p \neq q} (pq|qp)\hat{e}_{pqpq},\label{eq:Hd}
\end{equation}
with $\hat{E}_{pq}=\sum_{\sigma}a_{p\sigma}^\dagger a_{q\sigma}$
and $\hat{e}_{pqrs} = \sum_{\sigma,\tau} a^\dagger_{p\sigma}a^\dagger_{q\tau}a_{r\tau} a_{s\sigma}=E_{ps}E_{qr}-\delta_{qs}E_{pr}$. This Hamiltonian is spin-free,
however, it is only block-diagonal in the determinant space, since
it contains additional couplings for determinants with the same
spatial occupations\cite{p_DMRG}.
Thus, in this work, we instead used the Epstein-Nesbet (EN) form, which is equivalent
to neglecting the above off-diagonal couplings in the determinant space.
In the following, we will use $\hat{H}_d$ to denote the EN form,
which satisfies $\langle D_i|\hat{H}_d|D_j\rangle = \delta_{ij}\langle D_i|\hat{H}|D_j\rangle
\triangleq \delta_{ij}E_i$. This choice will result in a slight difference
in the computed $E_2$ compared with our previous p-DMRG, usually found to be less than 1m$E_h$ (vide post),
and the difference will gradually decreases as $M_0$ increases.

The EN partition is, of course, commonly used in SCI+PT\cite{huron1973iterative,buenker1974individualized,harrison1991approximating, schriber2016communication,tubman2016deterministic,liu2016ici,holmes_heat-bath_2016,sharma2017semistochastic,garniron2017hybrid}.
In this case, $Q[Q(\hat{H}_0-E_0)Q]^{-1}Q$ in \eqref{eq:e2init} can be simplified
to $Q(\hat{H}_d-E_0)^{-1}Q$ for $Q=\sum_{i}|D_i\rangle\langle D_i|$ and $[Q,\hat{H}_d]=0$, giving
$E_2=\sum_{i}-\frac{|\langle D_i|\hat{V}|\Psi^{(0)}\rangle|^2}{E_i-E_0}$
as a sum over determinants in the $Q$-space as mentioned above. However, in our case, $Q(\hat{H}_d-E_0)Q$ and
hence $Q[Q(\hat{H}_0-E_0)Q]^{-1}Q$ are not diagonal in the determinant space, due to the more complicated structure of $Q$ arising from the fact that the zeroth order MPS potentially can have nonzero overlap with every determinant. To derive a similar expresion for $E_2$, we first rewrite
\begin{eqnarray}
Q(\hat{H}_0-E_0)Q&=&Q(\hat{H}_d-E_0)Q\triangleq X-Y,\nonumber\\
X&=&(\hat{H}_d-E_0),\nonumber\\
Y&=&PX+XP-PXP.
\end{eqnarray}
In general, the operator $X=\hat{H}_d-E_0$ can be assumed to be invertible by
properly choosing $E_0$, and this important issue will be discussed later.
Then, using the relation $Q(X-Y)^{-1}Q=QX^{-1}(1-YX^{-1})^{-1}Q=
QX^{-1}Q+QX^{-1}YX^{-1}Q+QX^{-1}YX^{-1}YX^{-1}Q+\cdots$, by direct computation
we can find $Q(X-Y)^{-1}Q$ as
\begin{eqnarray}
Q(Q(\hat{H}_0-E_0)Q)^{-1}Q=
QX^{-1}Q
-\frac{QX^{-1}PX^{-1}Q}
{\bra{\Psi^{(0)}}X^{-1}\ket{\Psi^{(0)}}},\label{eq:Key}
\end{eqnarray}
which can be simply verified by multiplying the right hand side
with $Q(\hat{H}_0-E_0)Q$ to get the identity operator $Q$ in the $Q$-space.
The last term is the correction for the fact that $Q(\hat{H}_d-E_0)^{-1}P$
is in general nonzero, which clearly vanishes in the SCI+PT case as $[Q,\hat{H}_d]=0$.
Substituting Eq. \eqref{eq:Key} into Eq. \eqref{eq:e2init} for $E_2$
and invoking the resolution of identity $\sum_i|D_i\rangle\langle D_i|=1$, each term in $E_2$
can finally be formulated as a sum over determinants, viz.,
\begin{eqnarray}
E_2 &\triangleq& A+\frac{|B|^2}{C},\label{eq:E2}\\
A &\triangleq& -\langle\Psi^{(0)}|\hat{V}QX^{-1}Q\hat{V}|\Psi^{(0)}\rangle=-\sum_i \frac{|\langle\Psi^{(0)}|\hat{V}Q|D_i\rangle|^2}{E_i-E_0},\label{eq:A}\\
B &\triangleq& \bra{\Psi^{(0)}}\hat{V}QX^{-1}\ket{\Psi^{(0)}}=\sum_i \frac{\langle\Psi^{(0)}|\hat{V}Q|D_i\rangle\langle D_i|\Psi^{(0)}\rangle}{E_i-E_0},\label{eq:B}\\
C &\triangleq & \bra{\Psi^{(0)}}X^{-1}\ket{\Psi^{(0)}}=\sum_i \frac{|\langle\Psi^{(0)}|D_i\rangle|^2}{E_{i}-E_0}.\label{eq:C}
\end{eqnarray}
These formulae constitute the final working equations for stochastic p-DMRG.
Note that unlike in the case of SCI+PT, here the summation over $|D_i\rangle$
goes over \emph{all} the determinants.

In the practical evaluation of $A$, $B$, and $C$, instead of sampling $|D_i\rangle$ uniformly, we
use importance sampling to improve the efficiency.
For simplicity, we first discuss the evaluation of the term $C$ \eqref{eq:C}.
It can be evaluated as $C=\langle\frac{1}{E_{i}-E_0}\rangle_{P_i=|\langle\Psi^{(0)}|D_i\rangle|^2}$,
where the subscript indicates that the average of $\frac{1}{E_{i}-E_0}$ is taken with respect to the population of the determinants generated with the probability $P_i=|\langle\Psi^{(0)}|D_i\rangle|^2$.
To achieve this, we use an algorithm similar to that in Refs. \cite{METTS,stoudenmire_minimally_2010}.
Specifically, suppose $\ket{\Psi^{(0)}}$ is in the right canonical form,
\begin{equation}
  \ket{\Psi^{(0)}} = \sum_{n_1 n_2 \dots n_K} C^{n_1}[1]R^{n_2}[2]\dots R^{n_K}[K]\ket{n_1 n_2\dots n_K},
\end{equation}
where $K$ is the number of spatial orbitals, $C^{n_1}_{\alpha_1}[1]$ and $R^{n_K}_{\alpha_{K-1}}[K]$ are matrices, and $R^{n_k}_{\alpha_{k-1}\alpha_{k}}[k]$ are rank-3 tensors satisfying
the right canonical condition $\sum_{n_k\alpha_k}R^{n_k}_{\alpha_{k-1}'\alpha_{k}}[k]
R^{n_k}_{\alpha_{k-1}\alpha_{k}}[k]=\delta_{\alpha'_{k-1}\alpha_{k-1}}$,
then a determinant $|D_i\rangle=\ket{n_1 n_2\dots n_K}$ can be sampled
according to $P_i$  by a single sweep from left to right as follows:
The first occupation number $n_1\in\{|vac\rangle,|1_\beta\rangle,|1_\alpha\rangle,|1_\alpha1_\beta\rangle\}$ is generated according to $p_1(n_1)\triangleq\sum_{\alpha_1}|C^{n_1}_{\alpha_1}|^2$, which satisfies
$\sum_{n_1}p_1=1$ as $\ket{\Psi^{(0)}}$ is normalized.
Given $n_1$, at the second site, the generation probability for $n_2$
is defined as $p_2(n_2|n_1)\triangleq\sum_{\alpha_2}|(C^{n_1}R^{n_2})_{\alpha_2}|^2/N_2$.
Importantly, the normalization constant $N_2$ can be found as $N_2=\sum_{n_2}p_2(n_2|n_1)=p_1(n_1)$
due to the right canonical property.
Repeating this procedure recursively, the generation probability for $n_k$ given
$n_1\cdots n_{k-1}$ can be defined as $p_k(n_k|n_1\cdots n_{k-1})\triangleq
\sum_{\alpha_1}|(C^{n_1}R^{n_2}\cdots R^{n_k})_{\alpha_{k}}|^2/N_k$ with
$N_k=p_{k-1}(n_{k-1}|n_1\cdots n_{k-2})$.
Thus, after the occupation of the last site
is chosen, the total generation probability is $P(n_1\cdots n_K)=
p_1(n_1)p_2(n_2|n_1)\cdots p(n_k|n_1\cdots n_{K-1}) =
|(C^{n_1}R^{n_2}\cdots R^{n_K})|^2=|\langle\Psi^{(0)}|D_i\rangle|^2=P_i$, equal to our target distribution.
For a spin-adapted DMRG implementation\cite{sharma_spin-adapted_2012}, the determinants
can be generated similarly, but at each step the Clebsch-Gordon coefficient
$C^{S_kM_k}_{S_{k-1}M_{k-1}s_km_k}$ needs to be incorporated to map
each reduced tensor to the full one.

For term $A$ \eqref{eq:A}, it would be possible to employ the same strategy, if
$Q\hat{V}|\Psi^{(0)}\rangle$ can be expressed as an MPS faithfully.
However, since converting $Q\hat{V}|\Psi^{(0)}\rangle$ into an MPS exactly
would incur a bond dimension of $O(K^2M_0)$, this becomes prohibitive for large active spaces.
Thus, we first use a  bond dimension $M_1$ (that is small compared with $O(K^2M_0)$) to compress $Q\hat{V}|\Psi^{(0)}\rangle$
variationally as an MPS, i.e., $|\Phi\rangle\approx Q\hat{V}|\Psi^{(0)}\rangle$. This approximate state $|\Phi\rangle$ is then used to
define a generation probability for $|D_i\rangle$, $P_i=|\langle\Phi|D_i\rangle|^2$, for importance sampling, such that
\begin{eqnarray}
A= -\Big\langle \frac{|\langle\Psi^{(0)}|\hat{V}Q|D_i\rangle|^2}{P_i(E_i-E_0)}
\Big\rangle_{P_i=|\langle\Phi|D_i\rangle|^2}.\label{eq:Apgen}
\end{eqnarray}
In the case that $|\Phi\rangle$ is a good approximation to $Q\hat{V}|\Psi^{(0)}\rangle$, an approximate estimate for $A$ is just
\begin{eqnarray}
A\approx -\Big\langle \frac{1}{E_i-E_0}\Big\rangle_{P_i=|\langle\Phi|D_i\rangle|^2},\label{eq:Aapprox}
\end{eqnarray}
which becomes similar to the expression for term $C$ but with respect to a different distribution.
This approximation becomes exact in the limiting case that $|\Phi\rangle=Q\hat{V}|\Psi^{(0)}\rangle$.
It deserves to be noted that in Eq. \eqref{eq:Apgen}, there is a subtlety.
For the equality to hold, 
the set of determinants $\mathcal{S}\triangleq\{|D_i\rangle:\langle\Phi|D_i\rangle\ne 0\}$ interacting
with $|\Psi^{(0)}\rangle$ 
must include the set that interacts with $\langle \Psi^{(0)}|\hat{V}Q|D_i\rangle\ne 0$, otherwise
some contributions to $E_2$ will be missing. This is guaranteed for sufficiently large $M_1$
and fortunately, we found that to converge to chemical
accuracy ca. 1m$E_h$, the required $M_1$ can be much smaller than $O(K^2M_0)$ (vide post).
For the term $B$ \eqref{eq:B}, we simply evaluate it as
$B = \langle \frac{\langle\Psi^{(0)}|\hat{V}Q|D_i\rangle\langle D_i|\Psi^{(0)}\rangle}{P_i(E_i-E_0)}
\rangle_{P_i=|\langle\Phi|D_i\rangle|^2}$, since it turns out
to be less important in most cases, due to its small size and
the quadratic dependence of $E_2$ on it.

The above algorithm constitutes a stochastic algorithm for p-DMRG.
Its computational cost depends on two parts: the compression for $Q\hat{V}|\Psi^{(0)}\rangle$
on the right hand side of the first order equation \eqref{eq:1st},
which scales as $O(K^3M_1^2M_0)$ assuming $M_1\gg M_0$, and
the cost for the stochastic evaluation of $E_2$. The latter is dominated
by the number of samples $N_s$ times the cost for evaluating
the matrix elements $\langle\Psi^{(0)}|\hat{V}Q|D_i\rangle$.
We found the following scheme to be efficient at least for the
systems investigated in this work. By using the identity $\langle\Psi^{(0)}|\hat{V}Q|D_i\rangle=\langle\Psi^{(0)}|\hat{H}Q|D_i\rangle
=\langle\Psi^{(0)}|(H-E_{DMRG}^{(0)})|D_i\rangle
=\sum_j\langle\Psi^{(0)}|D_j\rangle\langle D_j|(H-E_{DMRG}^{(0)})|D_i\rangle$,
the evaluation is converted into $O(K^2N^2)$ evaluations of
overlaps between determinants and the zeroth order wavefunction $\langle\Psi^{(0)}|D_j\rangle$,
each of which scales as $O(KM^2_0)$. Thus, the total cost for the stochastic evaluation
is $O(N_sN^2K^3M^2_0)$, which is formally higher than $O(N_sK^3M^2_0)$ if the matrix element
$\langle\Psi^{(0)}|(H-E_{DMRG}^{(0)})|D_i\rangle$ is directly computed by considering
$|D_i\rangle$ as an MPS with bond dimension one. However, the efficiency of
our choice in practice may stem from the fact that in computing $\langle\Psi^{(0)}|D_j\rangle$,
the sparsity in the MPS tensors can be utilized, such that the actual
computational time is less than the direct evaluation of $\langle\Psi^{(0)}|(H-E_{DMRG}^{(0)})|D_i\rangle$.
A detailed comparison of these two choices for large orbital spaces will be presented in future.
The overall time of the stochastic step is usually found to be much smaller than
the compression step. Thus, the present stochastic algorithm
is more efficient than the previous p-DMRG algorithm, which required the iterative solution of the first order equation Eq. \eqref{eq:1st},
which scales as $O(K^2M_1^3+K^3M_1^2M_0)$, and with a much larger $M_1$ than required in the present algorithm.

We now demonstrate the accuracy and efficiency of the
stochastic algorithm in comparison with the previous deterministic p-DMRG algorithm for two
prototypical molecules: \ce{C2} at its equilibrium bond length 1.24253\AA\cite{C2_bondlength}
in the cc-pVDZ basis set\cite{doi:10.1063/1.462569}, and
\ce{Cr2} at equilibrium bond length 1.68\AA\cite{bondybey_electronic_1983} in the Ahlrichs' SV basis\cite{doi:10.1063/1.463096} and the cc-pVDZ-DK basis\cite{doi:10.1063/1.1998907}.
For \ce{C2} and \ce{Cr2} in the SV basis, all electrons were correlated,
resulting  orbital spaces with 12 electrons in 28 orbitals,
(12e, 28o) for \ce{C2}, and (48e,42o) for \ce{Cr2}, respectively.
For \ce{Cr2} in the cc-pVDZ-DK basis, the same orbital space as in our previous work\cite{p_DMRG} was employed, viz., (28e, 76o) with the $1s$, $2s$ and $2p$ frozen.
As in our previous work on deterministic p-DMRG\cite{p_DMRG},
the zeroth order energy is defined by an interpolation of two limits,
\begin{eqnarray}
E_0(\lambda) &=&(1-\lambda) E_{DMRG}^{(0)}+\lambda E_{d}^{(0)},\nonumber\\
E_{DMRG}^{(0)}&=&\langle\Psi^{(0)}|\hat{H}|\Psi^{(0)}\rangle,\nonumber\\
E_{d}^{(0)}&=&\langle\Psi^{(0)}|\hat{H}_d|\Psi^{(0)}\rangle.
\end{eqnarray}
For simplicity, in this work we only explored two cases, $\lambda=0$ and
$\lambda=\frac{1}{2}$, where in the former case $E_0(\lambda=0)=E_{DMRG}^{(0)}$,
and in the latter case $E_0(\lambda=\frac{1}{2})$ is an average $(E_{DMRG}^{(0)}+
E_{d}^{(0)})/2$, which in practice we found to deliver significantly better energies
for challenging systems, due to the more balanced treatment of the zeroth order state and the determinant-like perturbers\cite{p_DMRG}. The limit $\lambda=1$ was previously found to be prone to the intruder state problem and not considered here.

\begin{table}
\caption{
Total energy of \ce{C2} ($E$+75 in $E_h$) in the cc-pVDZ basis set (12 electrons in 28 orbitals)
obtained by DMRG and stochastic p-DMRG ($N_s$=36000)
with various $M_0$. The exact energy obtained with variational DMRG
with $M=4000$ is -75.731960$E_h$. Values in parentheses are
the statistical uncertainties.}
\label{tab:C2}\scriptsize
\begin{tabular}{ccccccc}
\hline \hline
       &    & \multicolumn{2}{c}{p-DMRG}   & & \multicolumn{2}{c}{stochastic p-DMRG} \\
\cline{3-4}\cline{6-7}
$M_0$  & DMRG  & $\lambda=0$ &  $\lambda=\frac{1}{2}$ && $\lambda=0$ &  $\lambda=\frac{1}{2}$ \\
\hline
50  &-0.708499 & -0.729676 & -0.731740 && -0.73032(8)  &-0.73253(9)	 \\
100 &-0.724500 & -0.731540 & -0.732029 && -0.73173(3)  &-0.73227(3)	 \\
200 &-0.729502 & -0.731845 & -0.731990 && -0.731919(8) &-0.732065(9) \\
400 &-0.731380 & -0.731939 & -0.731966 && -0.731947(3) &-0.731976(3) \\
\hline\hline
\end{tabular}
\end{table}

\begin{table}
\caption{Values of different components
in stochastic p-DMRG with $\lambda=0$
for \ce{C2} (in $E_h$) in the cc-pVDZ basis set:
$E_2$, $A$, $B$ and $C$ in Eqs. \eqref{eq:E2}-\eqref{eq:C}, respectively.}\label{tab:C2_components}
\scriptsize
\begin{tabular}{ccccc}
\hline\hline
$M_0$ & $A$  & $B$  & $C$ & $E_2$ \\
\hline
50 	&-0.02182(8) 	&-0.0009(1) 	& 2.521(5) 	& -0.02182(8) \\
100 &-0.00723(3)	&0.0015(1)		& 2.407(4) 	& -0.00723(3) \\
200 &-0.002418(8)	&0.00003(5)  	& 2.353(4)  & -0.002418(8)\\
400 &-0.000567(3)  &-0.000007(9)	& 2.343(4) 	& -0.000567(3)\\
\hline\hline
\end{tabular}
\end{table}

The results for \ce{C2} obtained with stochastic p-DMRG with $N_s=36000$ are listed in Table \ref{tab:C2} together with the p-DMRG results.
The orbitals were obtained by a CASSCF calculation with an active space CAS(6e,6o). It is seen that both p-DMRG and its stochastic variant improve on the variational
DMRG results at each $M_0$, and converge towards the exact value computed
by variational DMRG with $M=4000$ very quickly.
The $\lambda=\frac{1}{2}$ energies are better than those from
 $\lambda=0$, due to the reasons discussed earlier.
The bond dimension $M_1$ used for compression is chosen as 2000, which is sufficient
for Eq. \eqref{eq:Apgen} to hold. The total energies of
stochastic p-DMRG are seen to be very close to those from its
deterministic counterpart for each $\lambda$.
Neglecting off-diagonal couplings in Eq. \eqref{eq:Hd}, which is equivalent to employing
the EN partition, in the stochastic variant, leads to slightly lower energies.
This can be rationalized by the fact that after diagonalization of
$Q\hat{H}_dQ$ these couplings will
make the singlet perturbers have higher energies in the original p-DMRG
compared with $E_i$ for determinants,
and hence the original p-DMRG will have smaller second-order energies for singlet states.
It is seen that in all cases, the statistical errors are very small
with a moderate number of samples. The wall time at $M_0=400$
is about 1 minute for stochastic p-DMRG, which is
4 times faster than the deterministic one.

Next, the different components of $E_2$, viz., $A$, $B$ and $C$ in Eqs. \eqref{eq:A}-\eqref{eq:C},
are given in Table \ref{tab:C2_components}. The $B$ term is found to be very small,
such that the $|B|^2/C$ contribution is negligible for $E_2$ in this case.
However, since the calculation of $B$ and $C$ is relatively cheap compared with the evaluation of $A$,
we always include them in all our calculations.

The \ce{Cr2} dimer is a more difficult problem than \ce{C2}, and can be taken as a prototype of a
challenging molecule in quantum chemistry. The corresponding results are shown in Table
\ref{tab:Cr2}, with orbitals determined by a CASSCF calculation with CAS(12e,12o).
In the stochastic calculations, the bond dimension for compression is $M_1=8000$,
and the number of samples is $N_s=28000$ except for one column where 10 times
more samples were used to illustrate the convergence of the stochastic error.
From Table \ref{tab:Cr2}(a), we see that the usual variational DMRG converges very slowly,
and even with $M=8000$ the error is about 1m$E_h$ compared with the extrapolated result.
In contrast, the convergence to chemical accuracy using p-DMRG
and its stochastic variant is extremely fast using $\lambda=\frac{1}{2}$.
As shown in Table \ref{tab:Cr2}(b), for p-DMRG, chemical accuracy is achieved at $M_0=300$, while for stochastic p-DMRG it
is achieved at $M_0=200$ (as its energy is usually slightly lower
due to the same reasons as discussed for \ce{C2}).
The convergence with $\lambda=0$ is considerably slower compared with the $\lambda=\frac{1}{2}$ cases, which we found in our previous study
due to the unbalanced treatment of the zeroth order state
and determinant-like perturbers\cite{p_DMRG}. However,
it is still significantly cheaper than the variational DMRG
for the same level of accuracy due to the smaller bond dimension $M_0$.
All stochastic evaluations took less than 15 minutes (wall time on a 28-core node)
for a statistical error of less than 1m$E_h$, while the variational compression for $Q\hat{V}\ket{\Psi^{(0)}}$ took
about 30 minutes with $M_1=8000$.
Thus, in total, the stochastic algorithm is about 5 times faster than the deterministic algorithm, and further
parallelizes much better than the deterministic one, and uses much less memory and disk.

\begin{table}
  \caption{Total energy ($E$+2086 in $E_h$) of \ce{Cr2}
  in the Ahlrichs' SV basis (48 electrons in 42 orbitals) obtained by DMRG and stochastic p-DMRG.}\label{tab:Cr2}\scriptsize

\subcaption{DMRG energy ($E$ in $E_h$) and discarded weights ($w$) with a reversed sweep schedule.
The error bar for the extrapolated energy $(\infty)$ is estimated\cite{roberto} as 1/5
of the difference between the extrapolation energy and $M=8000$ energy.}
\begin{tabular}{ccccccc}
\hline\hline
$M$ & 500 & 1000 &  2000 & 4000 & 8000 & $\infty$\\
\hline
$w$ & 5.4$\times 10^{-4}$ & 2.1$\times 10^{-4}$ & 1.1$\times 10^{-4}$ & 5.8$\times 10^{-5}$ &  2.8$\times 10^{-5}$ &   \\
$E$ & -0.4955 &-0.5052 &-0.5108 &-0.5140 &-0.5158 &-0.5174(3) \\
\hline\hline
\end{tabular}

\subcaption{Stochastic p-DMRG energy ($N_s=28000$) obtained with 28000 samples.
All calculations were performed on a 28-core node.
Values in  parentheses are the statistical uncertainties.}
\begin{tabular}{cccccccccc}
\hline\hline
      &      & \multicolumn{2}{c}{p-DMRG} && \multicolumn{3}{c}{stochastic p-DMRG} \\
\cline{3-4}\cline{6-8}
$M_0$ & DMRG & $\lambda=0$  & $\lambda=\frac{1}{2}$ && $\lambda=0$  & $\lambda=\frac{1}{2}$ & $\lambda=\frac{1}{2}$$^a$ \\
\hline
100 &-0.4236 & -0.5000 & -0.5147 &&-0.5025(3)   &-0.5177(4)&-0.5182(1)     \\
200 &-0.4568 & -0.5068 & -0.5154 &&-0.5072(2)   &-0.5167(3)&-0.5167(1)     \\
300 &-0.4785 & -0.5107 & -0.5161 &&-0.5118(2)   &-0.5170(2)&-0.51726(8)    \\
400 &-0.4861 & -0.5123 & -0.5165 &&-0.5130(2)   &-0.5171(2)&-0.51728(7)    \\
500 &-0.4921 & -0.5136 & -0.5169 &&-0.5143(2)   &-0.5174(1)&-0.51751(6)    \\
\hline\hline
\multicolumn{8}{l}{$^a$ Results obtained with 10 times more samples.}\\
\end{tabular}
\end{table}

We found, in practice, that the bond dimension $M_1$ for compression does not need to be too large to achieve the chemical accuracy. In Table \ref{tab:Cr2_M1}, the dependence of the stochastic p-DMRG results on $M_1$ for $\ket{\Phi}$ are shown for $M_0=500$, $\lambda=0$, and $N_s=28000$. Using Eq. \eqref{eq:Apgen},
$M_1=4000$ is sufficient to converge to 1m$E_h$ accuracy.
For too small $M_1$, due to the problem of a poor importance sampling function,
which limits the set of the interacting determinants to be explored, the energy
is slightly higher than the converged ones with larger $M_1$.
It is interesting to see that the approximate estimate \eqref{eq:Aapprox}
requires much larger bond dimension, approximately by a factor of 4,
to achieve the same level of accuracy as Eq. \eqref{eq:Apgen},
although its statistical error is smaller with the same $N_s$.
Therefore, in practice this approximation is not very useful.

\begin{table}
  \caption{Dependence of the total energy ($E$+2086 in $E_h$) obtained by stochastic p-DMRG on the bond dimension $M_1$ used to compress $Q\hat{V}\ket{\Psi^{(0)}}$ as an MPS $\ket{\Phi}$.
  The results are illustrated with $M_0=500$ and $\lambda=0$.}\label{tab:Cr2_M1}\scriptsize
\begin{tabular}{ccc}
\hline\hline
$M_1$& $E_2$ from Eq. \eqref{eq:Apgen} & $E_2$ from Eq. \eqref{eq:Aapprox} \\
\hline
1000 &-0.5100(15)   & -0.50144(3) \\
2000 &-0.5122(14)   & -0.50735(6) \\
4000 &-0.5143(7)    & -0.51151(7) \\
6000 &-0.5148(7)    & -0.51272(8) \\
8000 &-0.5143(2)    & -0.51330(8) \\
\hline\hline
\end{tabular}
\end{table}

Finally, we compare the performance of the stochastic p-DMRG against the
deterministic p-DMRG for a larger problem, \ce{Cr2} in an orbital space
of (28e, 76o), in Table \ref{table:Cr2_dz}. For this system, the convergence
of the previous deterministic p-DMRG is very slow. Even with a sum of five MPS
with bond dimension $M_1=7500$ for the first-order wavefunction,
$\ket{\Psi^{(1)}}=\sum_{i=1}^{5} \ket{\Psi^{(1)}_i}$,
the deterministic p-DMRG energies differs from the extrapolated
energy ($M_1=\infty$) by 5-10m$E_h$ depending on $M_0$.
However, the energies provided by the stochastic p-DMRG with a given $M_1=10000$ and $N_s=80000$
are already very close to the extrapolated $M_1=\infty$ p-DMRG energies,
demonstrating the efficiency of the stochastic p-DMRG.
As we showed in our deterministic p-DMRG calculations~\cite{p_DMRG} one can further extrapolate in $M_0$ and $M_1$ to obtain
an estimate of the exact energy without any truncation error. In the future, we will explore
the possibility of such extrapolations with stochastic p-DMRG also.


\begin{table}
  \caption{Total energies ($E$+2099 in $E_h$) of \ce{Cr2} in the cc-pVDZ-DK basis set (28 electrons in 76 orbitals) computed by deterministic and stochastic p-DMRG with $\lambda=\frac{1}{2}$. 
  }\label{table:Cr2_dz}\scriptsize
  \begin{tabular}{ccccc}
    \hline\hline
    $M_0$ & 1000 & 2000 & 3000 & 4000 \\
    \hline
    DMRG     & -0.8346  &  -0.8617  & -0.8743   &   -0.8818  \\
    \hline
    p-DMRG$^a$ ($M_1=5\times7500$) & -0.9036 & -0.9035  &-0.9035   &   -0.9037 \\
    p-DMRG$^b$ ($M_1=\infty$)      & -0.9080 & -0.9109  &-0.9129   &   -0.9141 \\
    \hline
    stochastic p-DMRG & -0.905(2) & -0.909(2) &-0.909(1)  &   -0.911(1) \\
    \hline\hline
    \multicolumn{5}{l}{$^a$ Ref. \cite{p_DMRG}: energy for 
    $\ket{\Psi^{(1)}}=\sum_{i=1}^{5} \ket{\Psi^{(1)}_i}$ with $M_1=7500$.}\\
    \multicolumn{5}{l}{$^b$ Ref. \cite{p_DMRG}: extrapolated energy for $M_1=\infty$.}
  \end{tabular}
\end{table}

In summary, we have presented an efficient stochastic algorithm to overcome the previous bottleneck in p-DMRG\cite{p_DMRG} developed recently for problems with large active spaces. We demonstrated that, in
combination with a good choice of zeroth order Hamiltonian, the stochastic p-DMRG
algorithm can provide highly accurate total energies for challenging systems with
a significantly smaller amount of computational resources as compared with the deterministic p-DMRG, and both of
these are much cheaper than the original variational DMRG, in large orbital spaces
with a mix of correlation strengths.

{\it
{\bf Note:} During the finalization of this work, we became aware of a related
work in Ref. \cite{sharma2018stochastic}. This work assumed
that $E_2$ can be approximated by the term $A$ in Eq. \eqref{eq:A} and
used a different way to compute $E_2$ where $|\Psi^{(0)}\rangle$
was represented stochastically along the lines of the original stochastic
HCI+PT\cite{sharma2017semistochastic}. The accuracy of the final energies from
that algorithm should be comparable to the p-DMRG energies with the choice $\lambda=0$.
}

\section*{Acknowledgements}
This work was supported by the US National Science Foundation through CHE 1665333. Additional support
was provided by OAC 1657286. ZL is supported by the Simons Collaboration on the Many-Electron Problem.
GKC is a Simons Investigator in Physics.

\end{document}